\documentclass[
amsmath, amssymb, 
aps, prl, reprint,
twocolumn,
superscriptaddress,
floatfix,
showkeys,
xcolor
]{revtex4-1}

\usepackage[colorlinks, linkcolor = black, citecolor = black, filecolor = black, urlcolor = blue]{hyperref}

\usepackage{graphicx,braket,nicefrac}
\usepackage{siunitx,upgreek}
\usepackage{mathtools}

\begin{document}
\title{Quantum oscillator noise spectroscopy via displaced cat states}
\author{Alistair R. Milne}
\email{alistair.r.milne@gmail.com}
    \thanks{present address: Q-CTRL Pty Ltd, Sydney, NSW, 2006, Australia}
	\affiliation{
	ARC Centre of Excellence for Engineered Quantum Systems, The University of Sydney, School of Physics, NSW, 2006, Australia }
\author{Cornelius Hempel}
	\thanks{present address: Paul Scherrer Institute, 5232 Villigen PSI,
Switzerland}
	\affiliation{
	ARC Centre of Excellence for Engineered Quantum Systems, The University of Sydney, School of Physics, NSW, 2006, Australia }
\author{Li Li}
    \affiliation{Q-CTRL Pty Ltd, Sydney, NSW, 2006, Australia}
\author{Claire L. Edmunds}
    \affiliation{
	ARC Centre of Excellence for Engineered Quantum Systems, The University of Sydney, School of Physics, NSW, 2006, Australia }
\author{Harry J. Slatyer}
    \affiliation{Q-CTRL Pty Ltd, Sydney, NSW, 2006, Australia}
\author{Harrison Ball}
    \affiliation{Q-CTRL Pty Ltd, Sydney, NSW, 2006, Australia}
\author{Michael R. Hush}
    \affiliation{Q-CTRL Pty Ltd, Sydney, NSW, 2006, Australia}
\author{Michael J. Biercuk}
\email{michael.biercuk@sydney.edu.au}
	\affiliation{
	ARC Centre of Excellence for Engineered Quantum Systems, The University of Sydney, School of Physics, NSW, 2006, Australia }
     \affiliation{Q-CTRL Pty Ltd, Sydney, NSW, 2006, Australia}

\date{\today}     

\begin{abstract}
Quantum harmonic oscillators are central to many modern quantum technologies. We introduce a method to determine the frequency noise spectrum of oscillator modes through coupling them to a qubit with continuously driven qubit-state-dependent displacements. 
We reconstruct the noise spectrum using a series of different drive phase and amplitude modulation patterns in conjunction with a data-fusion routine based on convex-optimization. We apply the technique to the identification of intrinsic noise in the motional frequency of a single trapped ion with sensitivity to fluctuations at the sub-Hz level in a spectral range from quasi-DC up to 50~kHz.

\end{abstract}

\maketitle

Harmonic oscillators and their quantized excitations play a crucial role in many quantum systems relevant to quantum information processing. In trapped-ion~\cite{CiracZoller:1995,Sackett:2000,Leibfried:2003,Schmidt-Kaler:2003}, many superconducting~\cite{Cross:2015,Paik:2016, Song:2017,Blais:2020} and also electron spin-based~\cite{Borjans:2020} architectures they mediate qubit-qubit interactions in the form of motional modes carrying phonons or microwave resonators storing photons, respectively. Optomechanical coupling between the quantum motion of micro-resonators and photons may also be utilized as a universal transducer between stationary and optical qubits \cite{Aspelmeyer:2014}. Furthermore, in both trapped ion and superconducting platforms the oscillators themselves have recently been used to realize a logical qubit encoded in coherent superposition states~\cite{Fluehman2019,Campagne-Ibarcq.2020} allowing for an efficient implementation of a bosonic quantum error correcting code~\cite{Gottesman.2001}. In these systems, oscillator frequency fluctuations are often a limiting error source, degrading the mediated interactions and encoded bosonic states. Such fluctuations have been probed in trapped ions using coherent displacements \cite{McCormick:2019a,LX16909} and large superpositions of number states \cite{McCormick:2019b,LX16909}. Noise spectroscopy of the oscillator system can identify performance-limiting error sources, assess their relative weights, and inform appropriately tailored error-mitigation strategies through quantum control engineering~\cite{Ball:2021}. 

In this Letter, we experimentally demonstrate a method for the spectrally-resolved sensing of harmonic oscillator frequency fluctuations that is based on the interference of cat states \cite{Wineland:2013}. Deterministic generation of oscillator cat states has been demonstrated in a variety of systems \cite{Monroe:1996,Lo:2015,Vlastakis:2013,Hacker:2019,Leong:2020} and interference of the oscillator wavepackets has previously been applied to single photon detection~\cite{Hempel:2013}. Here, we apply a continuously-driven qubit-state-dependent displacement to the motional wavepacket of a single trapped ion, while inverting the drive phase at regular intervals to tune the protocol's peak sensitivity in frequency space. We combine this phase modulation with a shaped amplitude envelope defined by band-limited Slepian functions~\cite{Frey:2017,Norris:2018, Frey_2020} to suppress spurious signatures arising from spectral leakage at harmonics of the peak sensitivity.  The modulation pattern and pulse shape of the driving field translate to a filter transfer function in frequency space~\cite{Green:2013,Paz-Silva:2014} specific to each sensing sequence. Combining these with measurement results via a convex-optimization routine, we quantitatively reconstruct the noise spectrum. Our experiments using a single $^{171}$Yb$^+$ ion reveal previously unidentified narrowband spectral noise features on a radial mode which we probe with sensitivity to shifts in the mode frequency at the $\sim 0.5$ Hz level~\cite{Supp}.

\begin{figure*}[t!]
    \centering
    \includegraphics{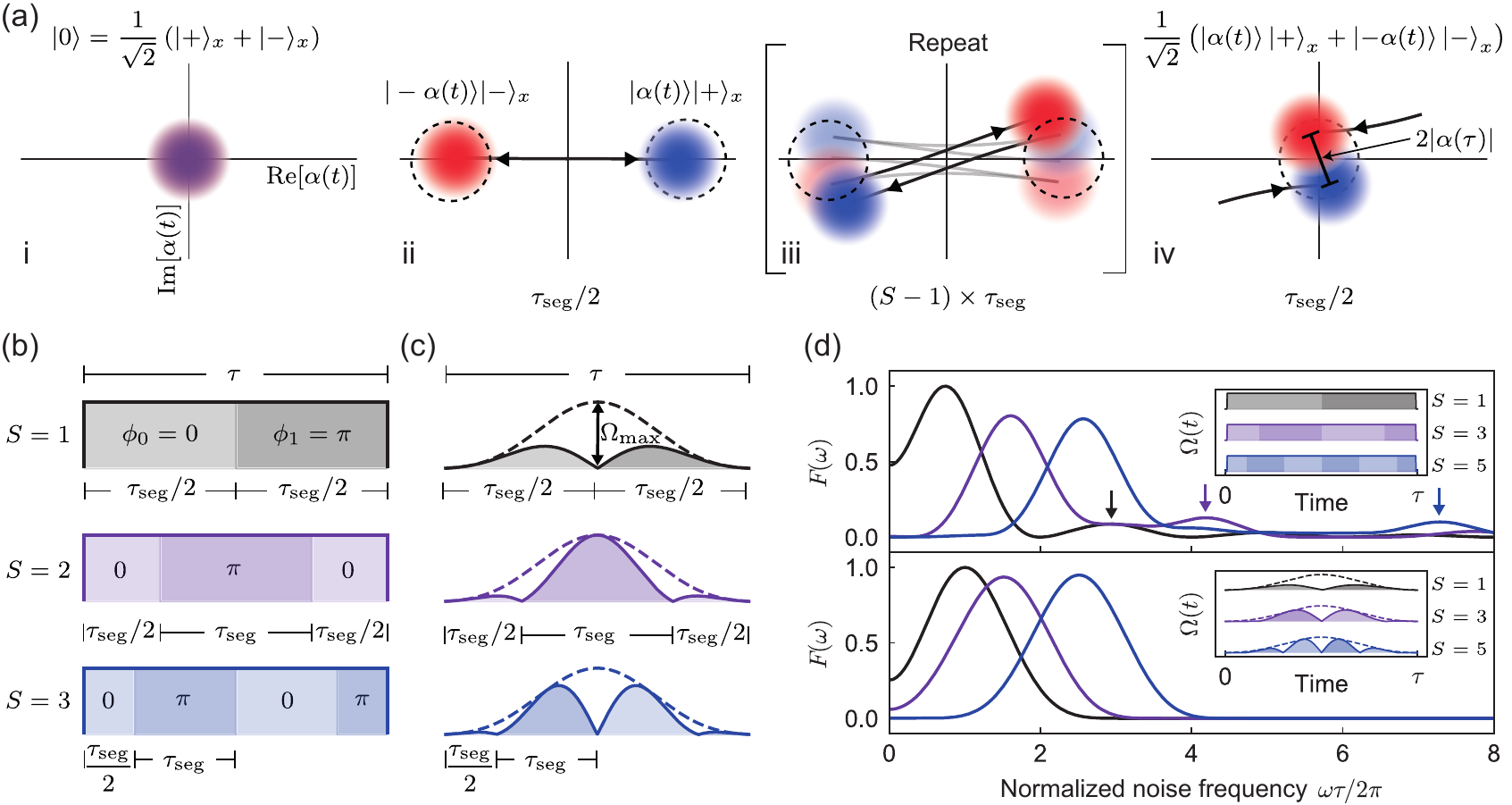}
    \caption{(a) Schematic illustration of the noise sensing sequences in oscillator phase space. (i) A qubit is prepared in $\ket{0}$ and the oscillator in its ground state (wavepacket centered at the origin). (ii) An initial displacement $\hat{D}(\hat{\alpha}(t))$ of duration $\tau_{\text{seg}}/2$ splits the wavepacket into two components associated with the $\ket{+}_x$ (blue) and $\ket{-}_x$ (red) internal qubit states, with the wavepackets following trajectories indicated by the black lines. (iii) The wavepackets are repeatedly displaced under $\pi$-phase inversion $(S-1)$~times for a duration of $\tau_{\text{seg}}$. The nominally straight trajectory in each segment curves under detuning noise and the wavepackets deviate from the nominal displacement at the conclusion of each segment (dashed circles). (iv) A final displacement of duration $\tau_{\text{seg}}/2$ recombines the wavepackets up to an accumulated differential displacement $2\vert\alpha(\tau)\vert$. (b,c) Schematic illustration of the fixed amplitude (square) and amplitude-modulated (Slepian) pulse profiles. The total sequence time $\tau$ and the number of phase shifts $S$ determine the segment duration $\tau_{\text{seg}}= \tau/S$. The coupling phase in each segment alternates between 0 and $\pi$, indicated by light and dark shading, respectively. For the amplitude-modulated sequences, a Slepian envelope (dashed line) with an underlying cosinusoidal modulation is applied to the maximum Rabi frequency $\Omega_{\text{max}}$. (d) Example filter functions $F(\omega)$ for the first three odd-$S$ square sequences (top), with arrows indicating harmonics, and for the equivalent Slepian-modulated sequences (bottom), where the harmonics have been suppressed. The filter functions are plotted against the dimensionless quantity $\omega\tau/2\pi$, which is the noise frequency $\omega$ normalized to the sequence duration $\tau$. Insets show the corresponding Rabi frequency $\Omega(t)$ and phase profiles.}
    \label{fig:schematic}
\end{figure*}

A bichromatic light field with frequency components symmetrically detuned from the red and blue motional sideband transitions couples the ion's internal state to the oscillator mode, via an interaction described by
\begin{equation}
	\hat{H}(t) = \frac{1}{2}\hbar\eta\Omega(t)\hat{\sigma}_{x} \left( e^{-i[\delta t + \phi(t)]}\hat{a}^{\dagger} + e^{i[\delta t + \phi(t)]}\hat{a}\right),
    \label{eq:MS_H}
\end{equation}

\noindent with Lamb-Dicke factor $\eta$, Rabi frequency $\Omega(t)$ and Pauli operator $\hat{\sigma}_{x}$ acting on the ion's internal state. The ion-oscillator coupling is captured by the creation and annihilation operators $\hat{a}^{\dagger},\,\hat{a}$ and the time-dependent exponential that includes the angular frequency difference $\delta$  (detuning) of the bichromatic field from the mode resonance, as well as the phase difference \mbox{$\phi(t) = (\phi_b(t) - \phi_r(t))/2$} between the two frequency components (blue and red).

The fundamental principle of the noise-sensing protocol is illustrated in Fig.~\ref{fig:schematic}(a), using a phase space representation of an oscillator co-rotating reference frame. Initially, the ion's internal state encoding a qubit is prepared in \mbox{$\ket{0}$} and the oscillator is brought close to its ground state (Fig.~\ref{fig:schematic}(a,i)). The unitary evolution of the system under Eq.~\eqref{eq:MS_H} enacts a qubit-state-dependent displacement of the motional wavepacket \cite{Roos:2008,Kirchmair:2009}, given by
\begin{equation}
    \hat{D}\left(\hat{\alpha}(t)\right) = \exp\left\{\hat{\sigma}_x\left( \alpha(t)\hat{a}^{\dagger} - \alpha(t)^{*}\hat{a}\right)\right\},
    \label{eq:displacement}
\end{equation}
where the displacement $\alpha(t)$ at time $\tau$ is given by \mbox{$\alpha(\tau) = \nicefrac{-i\eta}{2}\int_0^{\tau} \Omega(t)e^{-i\left[\delta t + \phi(t)\right]} dt$}. As the initial state $\ket{0}$ is a superposition in the displacement operator's $x$-eigenbasis, the application of Eq.~\eqref{eq:displacement} splits the ion wavepacket apart and creates an entangled state between the internal and oscillator degrees of freedom - a motional cat state, shown in Fig.~\ref{fig:schematic}(a,ii).

The sensing protocol is composed of an on-resonance ($\delta = 0$), continuously-driven state-dependent displacement with periodic discrete $\pi$ shifts of the coupling phase $\phi(t)$, inverting the direction in phase space. This structure repeatedly displaces the split wavepackets through the origin (Fig.~\ref{fig:schematic}(a,iii)). Finally, the wavepackets are brought back to the origin and, in the absence of noise, coherently re-interfere to restore the qubit state to $\ket{0}$.  The presence of motional mode frequency fluctuations, however, will result in curved displacements, gradually decreasing the overlap resulting in a separation of $2|\alpha(\tau)|$ at sequence end~(Fig.~\ref{fig:schematic}(a,iv)). This corresponds to residual qubit-oscillator entanglement and manifests as purity loss in projective measurements of the qubit. In the maximally mixed case of zero overlap between the wavepacket components, the probability $P_1$ of finding the $\vert1\rangle$ state reaches 0.5.

Each sequence in Fig.~\ref{fig:schematic}(b,c) is defined by the total duration $\tau$ and the number of phase shifts $S$. The total sensing pulse of length $\tau$ consists of $S+1$ segments indexed by $s\in\{0,S\}$, each with duration $\tau_s$ and coupling phase $\phi_s$ according to, 
\begin{equation*}
    \tau_s = 
\begin{cases}
   \tau_{\text{seg}}/2,& \text{if } s=0,S\\  
    \tau_{\text{seg}},              & \text{otherwise}
\end{cases}
\quad \text{and} \quad
    \phi_s = 
\begin{cases}
    0,& \text{if } s= \text{even}\\
    \pi,& \text{if } s= \text{odd}
\end{cases}
\end{equation*}
where $\tau_{\text{seg}} = \tau/S$, similar to a Carr-Purcell-Meiboom-Gill (CPMG) dynamical decoupling sequence \cite{Meiboom:1958}. The amplitude of the driving field may take the form of a flat-top ``square'' shape with a constant Rabi frequency \mbox{$\Omega(t) = \Omega_{\text{max}}$} (Fig.~\ref{fig:schematic}(b)). Alternatively, we may employ a smooth envelope determined by a modulated `Slepian' function (Fig.~\ref{fig:schematic}(c)), known to serve as a provably-optimal band-limited window in frequency space suppressing spectral leakage~\cite{Frey:2017, Norris:2018, Frey_2020}. Here, the modulated Rabi frequency, \mbox{$
    \Omega(t) = \Omega_{\text{max}}\Big| v_m^{(0)}(N,W)\cos\left\{\omega_{S}t\right\} \Big|,
    \label{eq:Omega_t}
$}
 consists of two components; the first is cosinusoidal with frequency \mbox{$\omega_S = 2\pi(\nicefrac{S}{2\tau})$}, matching the frequency of the alternating coupling phase. The second is an overall envelope defined by a zeroth-order Slepian $v_m^{(0)}(N,W)$, with sample number $N$ and bandwidth $W$~\cite{Supp}.  

In a measurement after sequence application, the expected value of $P_1$ (the sensor signal) may be expressed~\cite{Milne:2020} as the overlap integral of the noise power spectral density $S(\omega)$ and the sensing sequence's filter function $F(\omega)$ as
\begin{align}
    \mathbb{E}[P_1] &= \frac{1}{2\pi}\int_{-\infty}^{\infty}d\omega S(\omega)F(\omega) \quad\text{with}\\
     F(\omega) &= \sum_k T_k \left|\frac{2\pi\eta_k}{2} \int_0^{\tau} dt \Omega(t) e^{-i[(\delta_k-\omega)t + \phi(t)]}t\right|^2.
    \label{eq:FF}
\end{align}
The filter function, Eq.~\eqref{eq:FF}, describes the susceptibility of an operation to oscillator frequency noise at frequency $\omega$ summed over all oscillator modes $k$. Here, \mbox{$T_k = 2(\bar{n}_k + 1/2)$} incorporates the average initial phonon occupancy $\bar{n}_k$ for each mode (typically $\bar{n}_k \sim 0.2$ in our experiments). The first-order filter function is valid under the assumption that the residual oscillator displacement is small and that the noise is `weak' (see \cite{Supp}). In this work, we simplify the discussion by considering only a single mode, achieved by ensuring the oscillator mode frequencies are separated sufficiently such that driving a particular mode does not excite other modes. We focus exclusively on noise arising from fluctuations in the mode frequencies; it is assumed that amplitude noise on the drive field, to which the sequences are also susceptible, does not contain spectral components in the noise frequency domain of interest (see \cite{Supp} for further discussion).

For a fixed sequence duration $\tau$, increasing the number of phase shifts $S$ in either sequence type shifts the sensitivity of the filter function to higher frequencies (Fig.~\ref{fig:schematic}(d)). Peak sensitivity occurs near \mbox{$\omega_{\text{peak}} \approx 2 \pi \times S/2\tau$}, which applies to all Slepian-modulated cases with $S\geq 2$ and becomes increasingly accurate in the limit of large $S$ for the square sequences. Increasing the sequence duration $\tau$ narrows the filter bandwidth $\Delta\omega$ (defined as the full width at half maximum), with $\Delta\omega \sim 1/\tau$. Due to the abrupt inversion of the drive direction at each phase-shift, the frequency-space representation of square pulses exhibits higher order harmonics, which are suppressed under Slepian modulation. 

\begin{figure}
    \centering
    \includegraphics{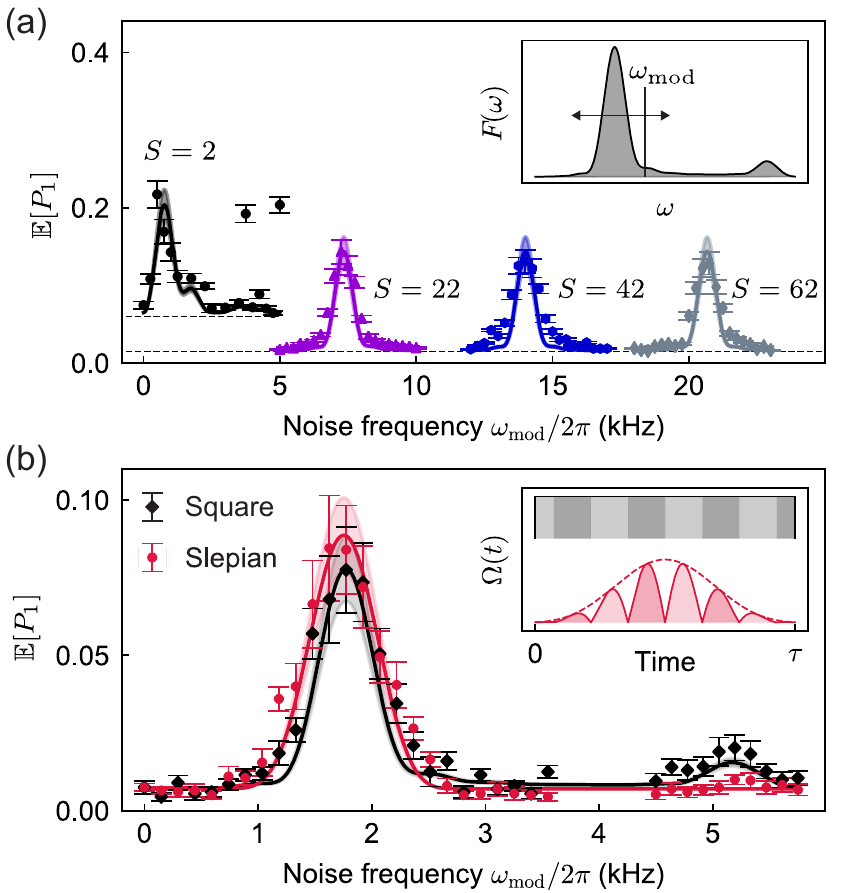}
    \caption{System-identification experiments. (a) Response to engineered single-tone noise with a depth of \mbox{$\beta_{\text{mod}} = 40$ Hz} for square sequences with \mbox{$\tau = 1.5$ ms} and $S = $~2, 22, 42 and 62. As illustrated in the inset, for each sequence the noise frequency $\omega_{\text{mod}}$ is scanned about the filter function peak. For the $S=2$ data, this frequency range includes the first harmonic of $F(\omega)$ (small bump in inset), for the other sequences this harmonic is not sampled. Experimental measurements of $\mathbb{E}[P_1]$ (markers) for each sequence are overlaid with filter function predictions (solid lines), including an additional frequency-independent offset (dashed horizontal lines). The additional offset for $S=2$ is likely due to the dominant intrinsic noise contributions in the low frequency regime (cf. Fig.~\ref{fig:intrinsic}), with the two spurious points potentially due to a transient increase in intrinsic noise. (b) Comparison of the response to single-tone noise for a square (black, diamonds) and Slepian-modulated (red, circles) sequence with \mbox{$\beta_{\text{mod}} = 65$ Hz}, \mbox{$\tau = 2$ ms} and \mbox{$S = 7$}. The Rabi frequency for the Slepian sequence (\mbox{$\Omega_{\text{max}} = 2\pi \times 6.2$ kHz}) is scaled relative to the square sequence (\mbox{$\Omega_{\text{max}} = 2\pi \times 2.6$ kHz}) in order to match peak sensitivity between the protocols. In both panels, error bars are the standard error of the mean (SEM) of the phase-samples averaged to give $\mathbb{E}[P_1]$ and shaded regions show uncertainty in the filter function prediction for a variation of $\bar{n} \pm 0.1$. The inset compares the Rabi frequency profile of the two sequences, with light and dark shading illustrating the alternating coupling phase.} 
    \label{fig:verification}
\end{figure}

We implement both approaches using a single $^{171}\text{Yb}^+$ ion confined in a linear Paul trap (similar to \cite{Guggemos:2017}), with motional frequencies $\omega_{x,y,z} \approx \{1.6,1.5,0.5\}$ MHz. A qubit is encoded on the \mbox{$\ket{F=0,m_F=0} \equiv \ket{0}$} and \mbox{$\ket{F=1,m_F=0} \equiv \ket{1}$} hyperfine ground state levels, split by $\sim 12.6$~GHz. Doppler cooling, state preparation and measurement are performed using a laser near 369.5~nm. Qubit and motional states are manipulated through  stimulated Raman transitions \cite{Hayes:2010} using two beams from a pulsed laser near 355~nm. We implement the state-dependent displacement by driving an acousto-optic modulator (AOM) in one of the beams with a two-tone radio-frequency signal, producing a bichromatic light-field that simultaneously drives the red and blue sideband transitions on resonance.

We first demonstrate the ability to produce a tuneable frequency response to motional frequency noise. A system identification procedure consisting of single-tone modulation at frequency $\omega_\textrm{mod}$ with magnitude $\beta_{\text{mod}}$ and phase $\phi_\textrm{mod}$ shifts the nominally resonant laser-frequency components symmetrically around the motional sideband frequencies in the form of $\beta_{\text{mod}}\sin\left(\omega_{\textrm{mod}}t + \phi_{\textrm{mod}}\right))$. In a given sensing sequence, we average the measured $P_1$ over different phase values $\phi_{\textrm{mod}}$ to obtain the expected value $\mathbb{E}[P_1]$, which we compare to theoretical predictions. For square pulses of different $S$ values we see good agreement between experiment and theory (Fig.~\ref{fig:verification}(a)). In these experiments we are, in principle, able to measure single-frequency signals above measurement-infidelity limits corresponding to detunings of $\sim10$ mHz, using sequences up to a duration of $\tau=32$~ms and $\Omega_\textrm{max}/2\pi = 30$~kHz \cite{Supp}. 

Comparing the response of the two kinds of sensing sequences provides direct evidence of harmonic-suppression in the filter function through Slepian amplitude-modulated waveforms. The data in Fig.~\ref{fig:verification}(b) shows that both sequences exhibit similar peak sensitivity at \mbox{$\omega_{\textrm{mod}}/2\pi \approx 1.8$~kHz}, with additional sensitivity due to spectral leakage at \mbox{$\omega_{\textrm{mod}}/2\pi \approx 5.2$~kHz} only present for the square sequence.

\begin{figure}[t!]
    \centering
    \includegraphics{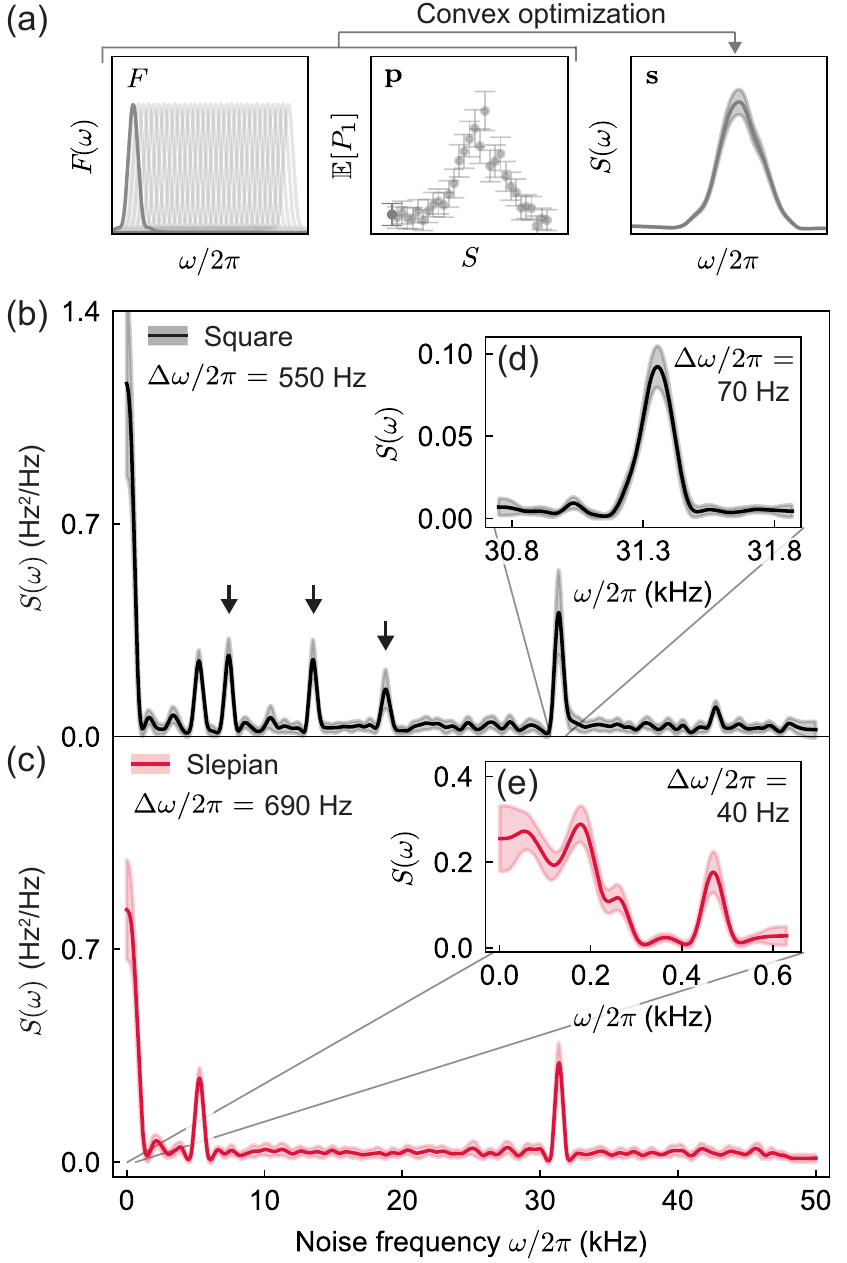}
    \caption{(a) Schematic of the noise reconstruction algorithm. A set of sequences described by filter functions $F$ produces a set of measurements ${\bf{p}}$. The noise spectrum ${\bf{s}}$ and associated uncertainty (shading) is determined by minimizing objective function  Eq.\eqref{eq:reconstruction}. (b,c) Reconstructed intrinsic noise power spectral density $S(\omega)$, comparing square (b) and Slepian (c) sequences with  $1 \leq S \leq 193$ and $\tau = 2$~ms with spectral resolution $\Delta\omega/2\pi$. The maximum Rabi frequencies used were {$\Omega_{\text{max}} = 2\pi \times 9$}~kHz for (b) and \mbox{$\Omega_{\text{max}} = 2\pi \times 20$}~kHz for (c). Features present only in the square reconstruction are indicated by black arrows. The insets (d,e) show reconstructions performed using higher spectral resolution sequences. The feature at $\sim 31.4$~kHz is probed with $\tau = 16$~ms, $986 \leq S \leq 1018$ and \mbox{$\Omega_{\text{max}} = 2\pi\times 0.9$}~kHz (d). The low frequency regime (e) is probed using Slepian sequences with $\tau = 32$~ms, $1 \leq S \leq 37$ and a reduced \mbox{$\Omega_{\text{max}}=2\pi\times 0.3$~kHz}. Reducing $\Omega_{\text{max}}$ ensures that the sensor response remains in the small-signal regime, prior to the onset of distortion due to higher-order filter terms, given by $\mathbb{E}[P_1]\lesssim0.1$~\cite{Green:2012,SoareNatPhys2014}.}
    \label{fig:intrinsic}
\end{figure}

Moving on from these validations, we require a technique to convert from measurements of $\mathbb{E}[P_1]$ in the presence of an unknown noise environment to a quantitative estimate of the noise power spectrum.  For a given set of filter functions $F$ and a vector of phase-averaged $\mathbb{E}[P_1]$ values, denoted as $\bf{p}$, the noise power spectrum $\bf{s}$ may be inferred via the relation \mbox{${\bf{p}} = F{\bf{s}}$} (Fig.~\ref{fig:intrinsic}(a)). We solve for $\bf{s}$ by employing an approach based on convex optimization~\cite{Ball:2021} used for the first time in experiment here.  In this framework the noise spectrum $\bf{s}$ is estimated by minimizing the objective function
\begin{equation}
    \text{min}_{\bf{s}}\left(||F{\bf{s}}-{\bf{p}}||_2^2 + \lambda \| D {\bf s} \|_2^2\right), {\bf{s}}\geq {\bf{0}}.
    \label{eq:reconstruction}
\end{equation}

\noindent The term $\lambda \| D {\bf s} \|_2^2$ is a regularization term where $D$ is the first order derivative operator (minimizing $\| D {\bf s} \|_2^2$ enforces smoothness in ${\bf{s}}$) and $\lambda$ is a hyperparameter tuned via a standard method to prevent under- or over-fitting \cite{Supp}; a strict-positivity requirement prevents overfitting-induced oscillations~\cite{Ball:2021}. This convex-optimization approach to spectrum estimation enables the use of arbitrary sets of measurements with no requirements on the underlying measurement probe structures unlike in dynamical decoupling based spectroscopy~\cite{alvarez_2011, Bylander:2011}. 

We use these techniques to detect and spectrally reconstruct intrinsic frequency noise on the ion’s radial motion. In Fig.~\ref{fig:intrinsic}(b,c), we sample noise in the band from zero to 50~kHz, comparing the reconstructions returned using both square and Slepian-modulated sequences.  We see a strong low-frequency signal that dominates the measured system performance, as well as a number of well defined noise features common to both data sets. The insets (Fig.~\ref{fig:intrinsic}(d,e)) show higher-spectral-resolution reconstructions of specific noise features, achieved by a combination of frequency shifting and increasing the pulse duration to narrow the filter bandwidth. The ability to arbitrarily shift the filter band and resolution - subject to hardware constraints - enables the noise spectrum to be probed in an iterative manner after identification of coarse spectral features~\cite{Norris:2018, Frey_2020}.  

The peak height $S(\omega_{\text{peak}})$ of the \textit{discrete} features present in the reconstructed spectra may be related to a motional frequency deviation $\pm \beta_{\text{dev}}$~Hz by taking into account the effect of the filter bandwidth $\Delta\omega$ using $\beta_{\text{dev}}\approx\sqrt{2S(\omega_{\text{peak}})(\Delta\omega/2\pi)}$, giving $\beta_{\text{dev}} \sim 10-20$~Hz for the observed features in Fig.~\ref{fig:intrinsic}(b,c). The sensitivity of the sequences employed in Fig.~\ref{fig:intrinsic}(b,c) is such that, in principle, the smallest detectable motional frequency deviations correspond to $\sim 7$~Hz for discrete and $\sim 0.3$~Hz for spectrally broad features. We have determined that the peaks near 5~kHz and 31~kHz likely arise from electromagnetic pickup in either the resonator stabilization circuit or the trap itself, having independently observed transient electromagnetic signals in the laboratory close to these frequencies. Further, frequency-resolved analysis of laser light shows no amplitude fluctuations commensurate with these features. We associate the additional spectral peaks present only in Fig.~\ref{fig:intrinsic}(b) with the sampling of out-of-band noise by the higher harmonics in the square-sequences' filter functions, or amplitude noise caused by rapid phase-transients in acousto-optic modulators which are suppressed by Slepian pulse modulation. See \cite{Supp} for further discussion on laser-amplitude noise, measurement sensitivity, and effects of sensor bandwidth.

In this work we have demonstrated that sequences of periodically inverted qubit-state-dependent oscillator wavepacket displacements provide a flexible means for performing noise spectroscopy on quantum oscillator modes via the creation of tuneable, band-limited filters for mode-frequency noise. The technique is readily-implementable in trapped-ion systems as it leverages the same interaction used to perform the ubiquitous M{\o}lmer-S{\o}rensen (MS) gate, enabling `in-situ' noise characterization with no additional hardware resources. We have employed this technique to sense noise on the radial motion of a single trapped ion in a spectral range from quasi-DC to 50~kHz, combining two distinct sensing waveforms with a convex optimization approach to spectrum estimation. In conjunction with previously reported  modulation and robustness protocols~\cite{Hayes:2012,Choi:2014,Haddadfarshi:2016,Manovitz:2017,Shapira:2018,Leung:2018a,Leung:2018b,Webb:2018,Zarantonello:2019,Lu:2019,Green:2015,Milne:2020,Landsman:2019,Shapira:2020,Grzesiak:2020}, our sensing technique provides a tool for designing gate operations in trapped-ion systems with robustness tailored to a specific noise environment.

\section*{Acknowledgements}
This work was partially supported by the Intelligence Advanced Research Projects Activity Grant No. W911NF-16-1-0070, the US Army Research Office Grant No. W911NF-14-1-0682, the Australian Research Council Centre of Excellence for Engineered Quantum Systems Grant No. CE170100009, and a private grant from H.\&A. Harley.
\newline

A.R.M. and C.H. contributed equally to this work.

\end{document}


\title{Supplemental material for ``Quantum oscillator noise spectroscopy via displaced cat states''}
\author{Alistair R. Milne}
\email{alistair.r.milne@gmail.com}
	\affiliation{
	ARC Centre of Excellence for Engineered Quantum Systems, The University of Sydney, School of Physics, NSW, 2006, Australia }
\author{Cornelius Hempel}
	\affiliation{
	ARC Centre of Excellence for Engineered Quantum Systems, The University of Sydney, School of Physics, NSW, 2006, Australia }
\author{Li Li}
    \affiliation{Q-CTRL Pty Ltd, Sydney, NSW, 2006, Australia}
\author{Claire L. Edmunds}
    \affiliation{
	ARC Centre of Excellence for Engineered Quantum Systems, The University of Sydney, School of Physics, NSW, 2006, Australia }
\author{Harry J. Slatyer}
    \affiliation{Q-CTRL Pty Ltd, Sydney, NSW, 2006, Australia}
\author{Harrison Ball}
    \affiliation{Q-CTRL Pty Ltd, Sydney, NSW, 2006, Australia}
\author{Michael R. Hush}
    \affiliation{Q-CTRL Pty Ltd, Sydney, NSW, 2006, Australia}
\author{Michael J. Biercuk}
\email{michael.biercuk@sydney.edu.au}
	\affiliation{
	ARC Centre of Excellence for Engineered Quantum Systems, The University of Sydney, School of Physics, NSW, 2006, Australia }
     \affiliation{Q-CTRL Pty Ltd, Sydney, NSW, 2006, Australia}

\date{\today}     

\maketitle

\subsection{Filter function prediction error scaling}

The filter function for trap frequency (motional mode) noise (Eq.~(4) in the main text) is only valid under certain assumptions about the character of the displacement sequences and noise~\cite{SoareNatPhys2014, Milne:2020}. Firstly, the residual displacement of the motional wavepacket at the conclusion of the sequence must be small, that is $\left|\alpha(\tau)\right|^2 \ll 1$. When the initial phonon occupancy of the oscillator is close to the ground state ($\bar{n} \approx 0$), the residual wavepacket displacement is approximately equal to $P_1$, thus this assumption may be equivalently stated as $P_1 \ll 1$. Secondly, the noise must be `weak' relative to the sensitivity of the displacement sequence, the condition for which is expressed as $\mathbb{E}[\epsilon(t)^2]\tau^2 \ll 1$. Here, $\epsilon(t)$ is a zero-mean error process that modulates the detuning from the average motional mode frequency as $\delta \rightarrow \delta + \epsilon(t)$. For sufficiently strong noise signals, the first-order filter function approximation is known to yield quantitative divergences between actual system response and prediction~\cite{Green:2012,SoareNatPhys2014}.

\begin{figure}[t!]
    \centering
    \includegraphics{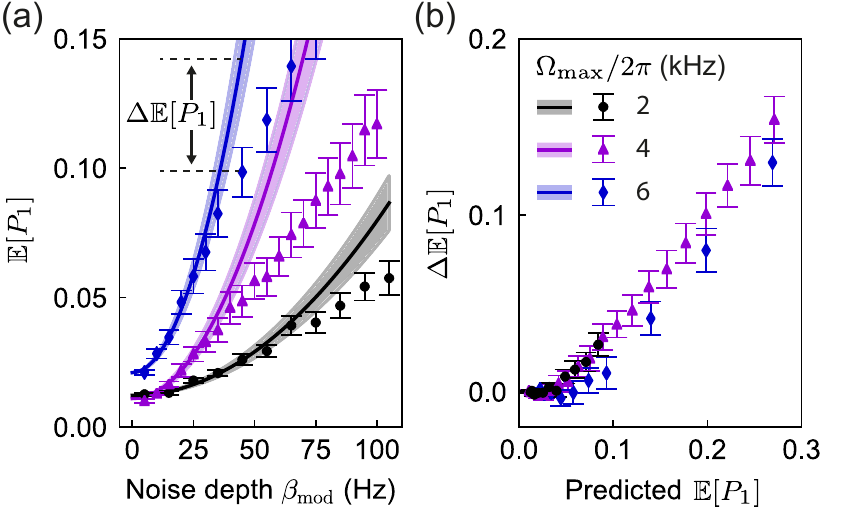}
    \caption{(a) Measured $\mathbb{E}[P_1]$ response to engineered single-tone noise with modulation depth $\pm \beta_{\text{mod}}$. The noise frequency $\omega_{\text{mod}}$ is set to match the peak frequency-sensitivity $\omega_{\text{peak}}$ of the sequence employed ($\tau = 1.5$ ms and $S =5$). Shown are the results for three different Rabi frequencies $\Omega_{\text{max}}/2\pi = 2$~kHz (black, circles), 4 kHz (purple, triangles) and 6 kHz (blue, diamonds).  Solid lines show filter function prediction for $\bar{n} = 0.9 \pm 0.1$ and markers are measured values, with error bars derived from quantum projection noise. (b) Difference $\Delta \mathbb{E}[P_1]$ between filter function prediction and measured $\mathbb{E}[P_1]$.}
    \label{fig:error}
\end{figure}

To investigate the conditions under which these assumptions remain valid, we engineer an effective trap frequency error in the same manner as the system identification experiments shown in Fig.~2 of the main text. In order to examine the system response and compare against filter-function predictions, we fix the noise frequency~$\omega_{\text{mod}}$ at the peak frequency-sensitivity~$\omega_{\text{peak}}$ of the sequence ($\tau = 1.5$~ms and $S=5$) and vary the magnitude of the applied error, $\pm\beta_{\text{mod}}$. Data shown in  Fig.~\ref{fig:error}(a) are the measured $\mathbb{E}[P_1]$ data for sequences implemented using three different Rabi frequencies, overlaid with the first order filter function predictions. 

In Fig.~\ref{fig:error}(b) we see that the difference between the predicted and measured signal, denoted as $\Delta \mathbb{E}[P_1]$, is negligible for predicted $\mathbb{E}[P_1] \lesssim 0.05$. For $\mathbb{E}[P_1] \gtrsim 0.05$, $\Delta \mathbb{E}[P_1]$ increases linearly with the predicted signal, indicating a quantitative overestimate of signal strength. In principle, calibrating the divergence between the predicted and measured signal enables larger values of $\mathbb{E}[P_1]$ to be included in a spectrum reconstruction, which increases the dynamic range of the sensing protocol by a factor of up to $\sim5-10\times$. In most practical noise spectroscopy applications, where trend and feature identification are more important than amplitude estimation within a factor of order unity, a user may simply ignore these divergences.

\subsection{Sequence sensitivity limits}

In the following, we examine the minimum motional frequency deviation detectable for a given sequence. We consider the case of a single tone noise feature as well as a noise density that is constant over the sequence filter function's band, $\Delta \omega$. For these two cases, we define the sensitivity of a given sequence as the motional frequency deviation which results in an $\mathbb{E}[P_1]$ signal of $P_{\text{min}}$, the minimum discernible signal above measurement baseline. The sensitivity is denoted $\beta_{\text{min}}$ when considering single-tone noise and $\beta_{\text{min,c}}$ for a constant noise density. In general, the sensitivity will depend on the total sequence duration $\tau$, the Rabi frequency $\Omega_{\text{max}}$ and whether the pulse profile is square or Slepian-modulated. The number of phase shifts $S$ does not affect the sensitivity, as modifying $S$ only adjusts the filter frequency-band without changing the filter magnitude.  

We start with the definition
\begin{align}
        \mathbb{E}[P_1] &=  \frac{1}{2\pi}\int_{-\infty}^\infty d\omega S(\omega)
	F(\omega).
	\label{eq:FF2}
\end{align}
For a single tone deviation of the motional frequency at frequency $\omega_{\text{dev}}$ with magnitude $\pm \beta_{\text{dev}}$~Hz, the noise density is defined as: 
\begin{align}
        S(\omega) = \frac{\beta_{\text{dev}}^2\pi^2}{2\pi}\left(\delta(\omega-\omega_{\text{dev}})+\delta(\omega+\omega_{\text{dev}})\right).
        \label{eq:sine}
\end{align}
Inserting this definition into Eq.~\eqref{eq:FF2} above, the expected signal for a single-tone noise feature under the application of a sequence with filter function $F(\omega)$ may be expressed as
\begin{align*}
        \mathbb{E}[P_1] = \frac{\beta_{\text{dev}}^2\pi^2}{4\pi^2}\left(F(\omega_{\text{dev}})+F(-\omega_{\text{dev}})\right).
\end{align*}
The filter functions for the sensing sequences we implement are symmetric about zero frequency, giving $F(\omega_{\text{dev}})=F(-\omega_{\text{dev}})$. Thus the expression for the expected signal can be simplified to
\begin{align}
        \mathbb{E}[P_1] &=  \frac{\beta_{\text{dev}}^2}{4}\left(F(\omega_{\text{dev}})+F(-\omega_{\text{dev}})\right) \nonumber \\
              &= \frac{\beta_{\text{dev}}^2}{2}F(\omega_{\text{dev}}). \label{eq:expected}
\end{align}
If we now assume the filter band is centered on the noise feature, that is \mbox{$\omega_{\text{dev}} = \omega_{\text{peak}}$}, and set $\mathbb{E}[P_1] = P_{\text{min}}$ (the minimum signal above the noise floor), the sensitivity for discrete signals $\beta_{\text{min}}$ is given by
\begin{align}
    \beta_{\min} = \sqrt{\frac{2P_{\text{min}}}{F(\omega_{\text{peak}})}}.
\end{align}

We now consider the case of a constant noise density across the filter band, characterized by the motional frequency deviation $\beta_c$. Here the noise density is defined as
\begin{align}
    S(\omega) = \frac{\beta_{c}^2\pi^2}{2\pi}
\end{align}
and the expected signal is  
\begin{align}
        \mathbb{E}[P_1] &=  \frac{\beta^2_{c}}{4}\int_{-\infty}^\infty F(\omega) d\omega. \nonumber
\end{align}
Rearranging, the expression for the sensitivity in the case of a constant noise density across the filter band $\beta_{\text{min,c}}$ is given by
\begin{align}
        \beta_{\text{min},\text{c}} = \sqrt{\frac{4P_{\text{min}}}{\int_{-\infty}^\infty F(\omega) d\omega}}.
\end{align}
As a specific example, consider the Slepian sequences used in the intrinsic noise reconstruction (Fig.~3(c) in the main text), with $\tau = 2$~ms and $\Omega = 2\pi \times 20$~kHz.  For a minimum sensor signal of $P_{\text{min}} = 0.01$, the sensitivities for this set of sequences are $\beta_{\text{min}} = 7.4$~Hz and $\beta_{\text{min,c}} = 0.28$~Hz. 

The best-case sensitivities given the parameter limitations of our system are $\beta_{\text{min}} = 8$~mHz and \mbox{$\beta_{\text{min,c}} = 1.3$~mHz} for a minimun sensor signal of \mbox{$P_{\text{min}} = 0.01$}. This is achieved using square sequences with \mbox{$\tau=32$~ms}, \mbox{$\Omega = 2\pi \times 30$~kHz} and assuming $\bar{n} = 0.2$.

\subsection{Effect of filter bandwidth on reconstructed feature height}

Here we analyze the effect of limited filter bandwidth on the resolution of the reconstructed spectrum, deriving the relation used to assign magnitudes to the discrete features present in the intrinsic noise spectra shown in the main text Fig.~3(b,c). Given a single-tone noise feature with frequency $\omega_{\text{dev}}$ (Eq.~\eqref{eq:sine}) sampled by a filter with $\omega_{\text{peak}} = \omega_{\text{dev}}$ and bandwidth $\Delta\omega$ (defined as the FWHM of the filter), we would like to relate the reconstructed feature height to the true frequency deviation, $\beta_{\text{dev}}$. We know that the overlap integral of the reconstructed spectrum with the filter $F(\omega)$ in a region centered about $\omega_{\text{peak}}$ should reproduce the measured $\mathbb{E}[P_1]$ signal, hence (using Eq.~\eqref{eq:expected}): 
\begin{align*}
    \frac{\beta_{\text{dev}}^2}{2}F(\omega_{\text{dev}}) = \frac{1}{2\pi}\int_{\omega_{\text{peak}} - \Delta\omega/2}^{\omega_{\text{peak}} + \Delta\omega/2}S(\omega)F(\omega)d\omega. 
\end{align*}
Setting $F(\omega_{\text{dev}}) = F(\omega_{\text{peak}})$ and approximating the integral on the right gives
\begin{align}
    \frac{\beta_{\text{dev}}^2}{2}F(\omega_\text{peak}) &\approx \frac{1}{2\pi}S(\omega_{\text{peak}})F(\omega_{\text{peak}})\Delta\omega.
\end{align}
From this we can relate the true deviation of the noise signal $\beta_{\text{dev}}$ to the reconstructed feature height $S(\omega_{\text{peak}})$ as
\begin{align}
    \beta_{\text{dev}} \approx \sqrt{2S(\omega_{\text{peak}})(\Delta\omega/2\pi)}. 
    \label{eq:conversion}
\end{align}

\begin{figure}[t!]
    \centering
    \includegraphics{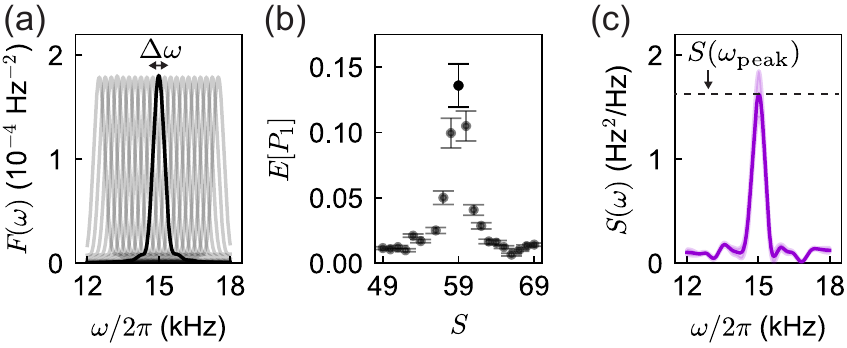}
    \caption{Reconstruction of an engineered single-tone noise feature with $\omega_{\text{mod}}/2\pi = 15$~kHz and $\beta_{\text{mod}} = 40$~Hz. (a) Filter functions corresponding to the square sequences implemented, with $\tau = 2$~ms, $49 \leq S \leq 69$, $\Omega_{\text{max}} = 6$~kHz and bandwidth $\Delta\omega/2\pi \approx 550$~Hz. The highlighted filter is centered on the frequency of the engineered noise. (b) $\mathbb{E}[P_1]$ measurements, with the data point corresponding to the central filter highlighted. (c) Reconstructed spectrum with $S(\omega_{\text{peak}}) \approx 1.6$ Hz$^2/$Hz.}
    \label{fig:delta}
\end{figure}

We experimentally verify Eq.~\eqref{eq:conversion} by reconstructing the spectrum of an effective single-tone trap frequency error (\mbox{$\omega_{\text{mod}}/2\pi = 15$~kHz} and $\beta_{\text{mod}}=40$~Hz), engineered via modulation of the laser detuning. The noise is sampled using a series of square sequences with corresponding filter functions centered about the noise frequency (Fig.~\ref{fig:delta}(a)), producing the phase-averaged $\mathbb{E}[P_1]$ measurements shown in Fig.~\ref{fig:delta}(b). The reconstructed feature (Fig.~\ref{fig:delta}(c)) has a peak height of \mbox{$S(\omega_{\text{peak}}) \approx 1.6 \pm 0.2$~Hz$^2/$Hz} and a width which is limited by the bandwidth of the sensing filter functions (\mbox{$\Delta\omega/2\pi \approx 550$~Hz}). Using Eq.~\eqref{eq:conversion}, these parameters give \mbox{$\beta_{\text{dev}} \approx 42 \pm 3$~Hz}, which closely matches the true magnitude of the engineered signal.  

\subsection{Estimating ion temperature}

A critical parameter in the filter function (Eq.~(4) in the main text) is the initial mean phonon occupancy, $\bar{n}$. Before performing a series of measurements, we determine this parameter by performing Rabi flopping on the blue sideband and performing a fit to the expression,

\begin{equation}
    P_1 = \frac{1}{\bar{n}+1}\sum_{n=0}^{n_{\textrm{max}}}\frac{p_n(\bar{n})}{2}\left(1-e^{-\gamma t}\cos\left(\Omega(n)t\right)\right),
\end{equation}

where $\bar{n}$, the base Rabi frequency $\Omega_0$ and the decay envelope $\gamma$ are simultaneous fit parameters. Here, the frequency components of the Rabi oscillations \mbox{$\Omega(n) = \Omega_0\eta\sqrt{n+1}$} are weighted by a thermal distribution \mbox{$p_n(\bar{n}) = \left(\nicefrac{\bar{n}}{\bar{n}+1}\right)^n$}. The sum over phonon states $n$ is typically truncated at $n_{\text{max}} = 100$. The decay envelope $\gamma$ accounts for incoherent damping caused, e.g., by fast intensity fluctuations in the light field, that is  not related to the modulation resulting from the underlying phonon statistics. 

Note that for the system identification experiments shown in Fig.~2(a) of the main text, this procedure was not performed. Instead, the filter function predictions for the $\mathbb{E}[P_1]$ signal are calculated using a typical value of $\bar{n} = 0.2 \pm 0.1$.

\subsection{Slepian amplitude-modulated sequences}

For Slepian-modulated sequences, $\tau_s$ and $\phi_s$ are defined the same as for the fixed-amplitude sequences. As described in the main text (and schematically illustrated in main text Fig.~1(c)), the modulation of the Rabi frequency consists of two components, the first of which is a cosinusoidal modulation of the form \mbox{$|\cos(\omega_{S}t)|$}, with $\omega_S = 2\pi(\nicefrac{S}{2\tau})$. The frequency $\omega_S$ is commensurate with the alternating coupling phase and thus the effect of this modulation is to apply a sinusoidal envelope to each phase segment. The second component is an overall envelope defined by a discrete prolate spheroidal sequence (DPSS, commonly referred to as a `Slepian' sequence), which is a discrete time function that maximises spectral concentration in a defined frequency band  \cite{Frey:2017, Norris:2018, Frey_2020}. We define \mbox{$v_m^{(k)}(N,W)$} as the $k\textsuperscript{th}$-order Slepian sequence consisting of $N$ points indexed by $m\in\{0,N-1\}$ and characterised by a sampling time $\Delta t$, such that $\tau = N\Delta t$. The discrete Fourier transform of \mbox{$v_m^{(k)}(N,W)$} is spectrally concentrated in \mbox{$[-2\pi W/\Delta t,2\pi W/\Delta t]$}, where $W\in(0,1/2)$ is the half-bandwidth parameter. The sequence is scaled such that \mbox{$\text{min}\{v_m^{(k)}(N,W)\} = 0$} and \mbox{$\text{max}\{v_m^{(k)}(N,W)\} = 1$}. In practice, the number of points $N$ is chosen such that the resulting profile is sufficiently smooth; the half-bandwidth parameter is then set using the relation $NW = k + 1$. Combining the Slepian and cosinusoidal modulation produces an overall Rabi frequency profile of the form  
\begin{equation}
    \Omega(t) = \Omega_{\text{max}}\Big| v_m^{(0)}(N,W)\cos\left\{\omega_{S}t\right\} \Big|.
    \label{eq:Omega_t}
\end{equation}

For amplitude-modulated sequences with an even number of phase shifts, the phase space trajectories do not nominally return to the origin at the conclusion of the operation. This is due to a difference in the total pulse area between segments with $\phi = 0$ and $\phi = \pi$. The effect is most pronounced at low numbers of phase shifts where the difference in pulse area is the largest. Because of this imbalance, only odd-$S$ amplitude-modulated sequences are suitable for noise sensing for small $S$ (low band-shift modulation frequencies). As $S$ is increased, the effect becomes less significant and both even-$S$ and odd-$S$ sequences may be used.

\subsection{Spectrum reconstruction by convex optimization}

By discretization of Eq.~(3) in the main text, solving the noise reconstruction problem is to find the solution ${\bf s}$ to the linear equation:
\begin{align}
\label{eq_linear_equation}
    F {\bf s} = {\bf p} \, \quad {\text {\textit{s.t.}}} \quad {\bf s} \geq {\bf 0}.
\end{align}
Here, $F$ is the filter function matrix; vector $\bf p$ is the measured $\mathbb{E}[P_1]$ values (infidelities); and vector $\bf s$ is the noise density. The noise frequency domain $[\omega_{\text{min}},\omega_{\text{max}}]$ is discretized into $n$ samples indexed by $i \in \{1,...,n\}$ such that
\begin{equation}
    \omega_i = \omega_{\text{min}} + (i-1)\Delta\omega_s
\end{equation}
with the sampling interval $\Delta\omega_s$ given by
\begin{equation}
    \Delta\omega_s = \frac{\omega_{\text{max}} - \omega_{\text{min}}}{n-1}.
\end{equation}
Given this discretization, for a set of $m$ measurements (corresponding to $m$ different sensing sequences) the filter function matrix $F$ is defined as
\begin{align}
    F =  \frac{\Delta\omega_s}{2\pi}\begin{bmatrix}
F_1^1 & F_2^1 & \cdots & F_{n-1}^{1} & F_n^1\\
F_1^2 & F_2^2 & \cdots & F_{n-1}^2 & F_n^2\\
\vdots  &  \vdots & \ddots & \vdots & \vdots\\
 F_1^m  & F_2^m  & \cdots & F_{n-1}^m & F_n^m \\
\end{bmatrix}
\end{align}
where the element $F^q_i$ is the value of the filter function corresponding to the $q$\textsuperscript{th} sequence at frequency $\omega_i$, with \mbox{$q \in \{1,...,m\}$}. In a similar manner, ${\bf s}$ and ${\bf p}$ are defined as:
\begin{align}
    {\bf s} =  \begin{bmatrix}
s_1 & \\
s_2 & \\
\vdots  & \\
s_n  &\\
\end{bmatrix} \quad \text{and} \quad
    {\bf p} =  \begin{bmatrix}
p_1 & \\
p_2 & \\
\vdots  & \\
p_m  &\\
\end{bmatrix},
\end{align}
where $s_i$ is the noise density at frequency $\omega_i$ and $p_q$ is the $\mathbb{E}[P_1]$ measurement for the $q$\textsuperscript{th} sequence. Typically, one would have $m \ll n$, which means there are in fact infinitely many solutions to Eq.~\eqref{eq_linear_equation}. This problem is generically an ill-posed inverse-problem~\cite{Tarantola:2005}.

Given this circumstance, the solution is not unique and usually not stable (sensitive to the changes in $F$). The key is then to find one or a set of solutions that will be reasonable to use based on some \emph{prior} information about the system. In this case, for instance, we know that the reconstructed noise PSD must be non-negative.  Introducing such a constraint into more general machine-learning approaches is not trivial and motivates an alternate approach.

Here, instead of directly finding a solution to Eq.~\eqref{eq_linear_equation}, we reformulate the problem with the prior information as an objective function. By minimizing that objective function, the optimal solution should represent the best solution as follows:
\begin{align}
\label{eq_cp_general}
    {\rm min}_{\bf s} ( \| F{\bf s} - {\bf p} \|_2^2 + \lambda R({\bf s}) ) \, \quad {\text {\textit{s.t.}}} \quad {\bf s} \geq {\bf 0} \; .
\end{align}
Here $\| \bullet \|_2$ means the Euclidean norm; $\lambda$ is a hyperparameter with positive value. $R({\bf s})$ is known as the regularization term and the choice of this term depends on specific inverse problems. Often $R({\bf s})$ is chosen as a convex function such that both objective function and constraint in Eq.~\eqref{eq_cp_general} are convex, permitting efficient numerical solutions.

In our work, the regularization term is chosen as $R({\bf s}) = \| D {\bf s} \|_2^2$, where $D$ is the first order derivative operator defined as:
\begin{align}
    D =  \begin{bmatrix}
-1 & 1 &   & & \\
   & . & . & & \\
   &   & . & . &\\
   &   &   & -1 & 1 \\
\end{bmatrix}_{(n- 1) \times n}
\end{align}
Minimizing $\| D {\bf s} \|^2_2$ is the equivalent to minimizing the difference between adjacent elements of ${\bf s}$, meaning we are expecting ${\bf s}$ is \emph{smooth} in the parameter space~\cite{Tikhonov:1963}. Given limited measurements, this smoothness assumption ensures a non-parametric approach to spectral estimation primarily captures broad trends (such as a $1/\omega$ background) rather than overfitting due to small and potentially itinerant features.

The hyperparameter used in the optimization process represents an effective weight for the inclusion of our regularization term.  Appropriate selection is a complex process in general; we adopt a strategy called cross-validation which is widely used in machine learning \cite{Hastie}. 

We first divide experimental data $(F, {\bf p})$ into two sections: $(F_1, {\bf p}_1)$ and $(F_2, {\bf p}_2)$. For a given hyperparameter $\lambda_j$ from the candidate set $\{\lambda_1, ..., \lambda_c\}$,
using $(F_1, {\bf p}_1)$ as the training data and $(F_2, {\bf p}_2)$ as the testing data, we find the solution ${\bf s}_j$ by solving the optimization problem with respect the training data set:
\begin{align}
\label{eq-train}
    {\rm min}_{\bf s} ( \| F_1 {\bf s} - {\bf p}_1 \|_2^2 + \lambda_j \| D {\bf s} \|_2^2 )\;,
\end{align}
and verify the solution using the testing data by calculating the testing error as:
\begin{align}
    e_j^1 = \|F_2 {\bf s}_j - {\bf p}_2 \|_2;.
\end{align}
Then we swap these two data sets, {\em i.e.}, using $(F_2 , {\bf p}_2)$ as the training data and $(F_1 , {\bf p}_1)$ as the testing data, repeat the above steps and find the testing error $e_j^2$. The optimal hyperparameter is chosen as the one to minimize the average testing error $(e_j^1 + e_j^2 )/ 2$.  The adoption of this technique as the basis for hyperparameter selection mitigates issues with manual tuning of the algorithm and is applied to all data in this work.

In our circumstance we construct the training and testing data sets in order to ensure that both data sets span the entire range of frequencies under test. Even rows of the filter function matrix and corresponding elements of the infidelity vector are used as $(F_1, {\bf p}_1)$ and the remainder are assigned as $(F_2, {\bf p}_2)$. Furthermore, in the training step, to capture the measurement uncertainties and SPAM error in the experiment, we generate a set of $\{{\bf p}_1^1, ..., {\bf p}_1^l\}$ by resampling from a Gaussian distribution $N({\bf p}_1, {\Delta}_1)$ where $\Delta_1$ are the measurements uncertainties.  We then run reconstruction as described by Eq.~\eqref{eq-train} with the ensemble of sampled infidelities. That is, for each ${{\bf p}_1^k}$ we have a reconstructed ${\bf s}_{jk}$ as per Eq.~\eqref{eq-train} and a corresponding testing error $e_{jk}^1$. The average testing error is then calculated as ${\bar{e}_j^1} = \frac{1}{l} \sum_{k=1}^l e_{jk}^1$. The same process is then utilized to calculate ${\bar{e}_j^2}$.

\subsection{Intrinsic noise reconstruction data}

\begin{figure}
    \centering
    \includegraphics{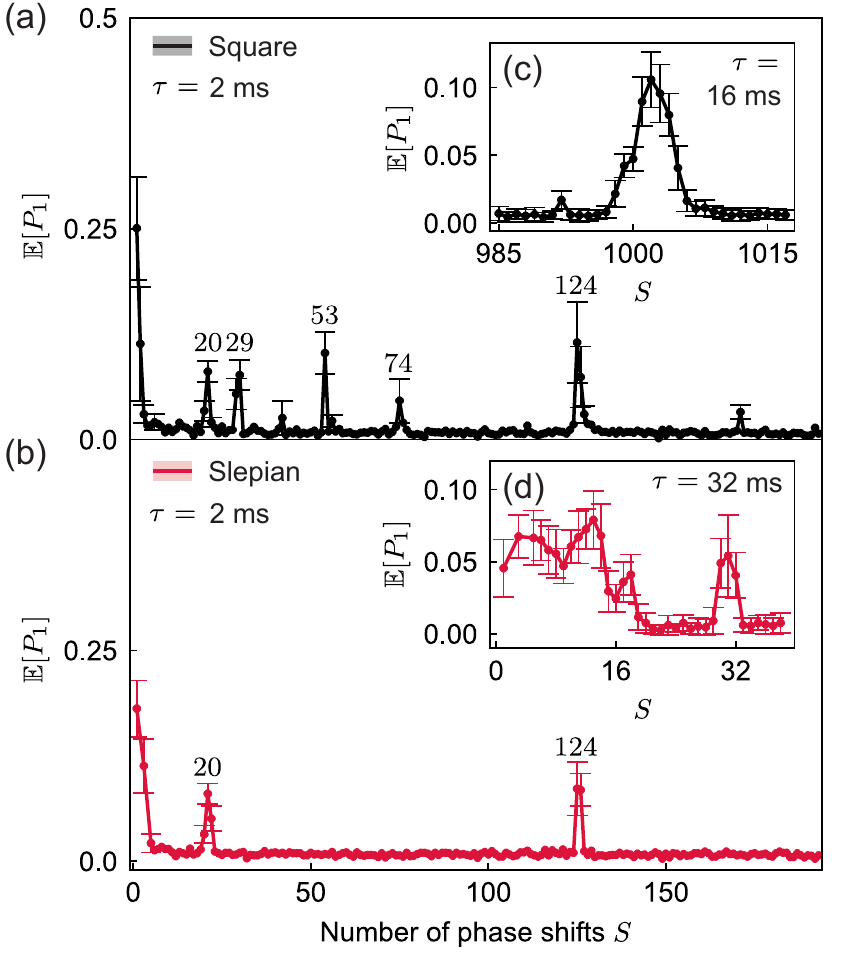}
    \caption{$\mathbb{E}[P_1]$ measurements used to perform the intrinsic noise reconstruction. (a,b) Results for square (a) and Slepian (b) sequences with $\tau = 2$~ms, $1 \leq S \leq 193$, probing the noise in the range \mbox{$\sim$~0 to 50 kHz}. Maximum Rabi frequencies are {$\Omega_{\text{max}} = 2\pi \times 9$}~kHz (square) and \mbox{$\Omega_{\text{max}} = 2\pi \times 20$}~kHz (Slepian). $\mathbb{E}[P_1]$ signal significantly above baseline is annotated with the $S$-value of the underlying sequence. (c) Results for square sequences with $\tau = 16$~ms, $986 \leq S \leq 1018$ and \mbox{$\Omega_{\text{max}} = 2\pi\times 0.9$}~kHz, probing noise in the range \mbox{$\sim$~30.8 to 31.8 kHz}. (d) Results for Slepian sequences with $\tau = 32$~ms, $1 \leq S \leq 37$ and \mbox{$\Omega_{\text{max}} = 2\pi\times 0.3$}~kHz, probing noise in the range \mbox{$\sim$ 0 to 0.6 kHz}.}
    \label{fig:data}
\end{figure}

Shown in Fig.~\ref{fig:data} are the $\mathbb{E}[P_1]$ measurements used to perform the intrinsic noise reconstruction (Fig.~3 in the main text). Each data point in the figure corresponds to a particular sensing sequence and is the average of $M$ consecutive $P_1$ measurements, each consisting of $r$ sequence repetitions (typical values are $M=10$ and $r=200$). Breaking up the measurement in this manner --as opposed to simply taking $M \times r$ repetitions in one shot-- introduces delays due to software timing between measurement points, randomizing the timing of the sensing protocol relative to the phase of the noise process being sensed. The error bars assigned to each data point are the standard deviation $\Delta$ of the $M$ measurements; these values are used in resampling step of the cross-validation procedure for hyperparameter selection.  

Fig.~\ref{fig:data}(a,b) compares the $\mathbb{E}[P_1]$ response using square and Slepian sequences of equal duration $\tau$, number of phase shifts $S$ and sensitivity $\beta_{\text{min}}$, with the sensitivity tuned via scaling of the Rabi frequencies $\Omega_{\text{max}}$. For both sequences types, there is a strong signal at low phase shifts as well as discrete features at higher $S$-values, with the square sequences showing additional features not present in the Slepian case. 

The data in Fig.~\ref{fig:data}(c) is a probe of the noise feature at $S=124$ in Fig.~\ref{fig:data}(a,b) performed with higher spectral resolution. Increased resolution is achieved by modifying the pulse duration from $\tau = 2$~ms to $16$~ms which narrows the filter bandwidth by $\sim 8 \times$. The number of phase shifts is also increased by the same factor from 124 to $\sim 1000$, ensuring $\omega_{\text{peak}}$ is centered on the noise frequency. In a similar manner, the data in Fig.~\ref{fig:data}(d) is the result of probing the low frequency regime with narrow bandwidth $\tau = 32$~ms Slepian sequences. 

\subsection{Out-of-loop noise characterisation}

\begin{figure}[t!]
    \centering
    \includegraphics{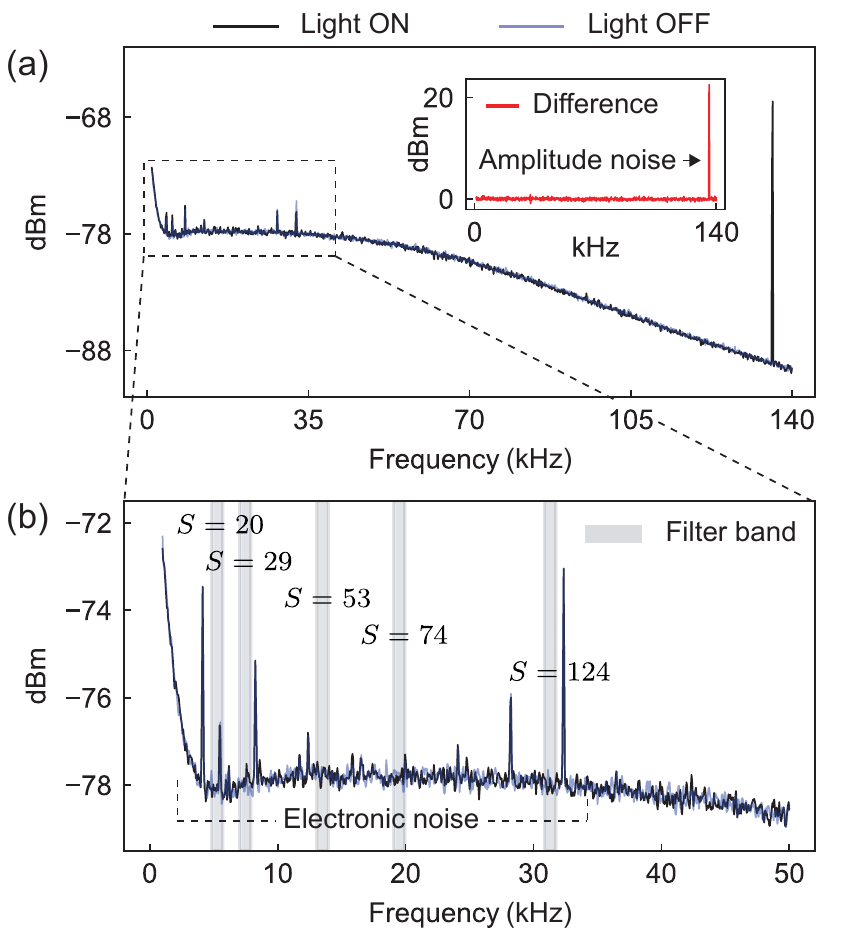}
    \caption{(a) Spectrum of the photodiode signal used to monitor the amplitude of the bichromatic 355~nm beam, comparing the case of light on (black) and off (blue). The inset shows the difference of the two traces (red). (b) A zoom-in of the 1-50 kHz region. The vertical gray bars indicate the main-band frequency-sensitivity of sequences which resulted in intrinsic noise signal in Fig.~3(b,c).}
    \label{fig:amplitude}
\end{figure}

In addition to motional frequency noise, the sensing sequences are also sensitive to laser amplitude noise, with the peak frequency-sensitivity of the amplitude-noise filter functions, $\omega_{\text{peak}}$, matching that of the motional frequency filter functions (for details on the amplitude noise filter functions, see \cite{Green:2015}). As such, it is important to characterize the laser amplitude noise spectrum to disambiguate signals arising from the two noise sources. 

In the intrinsic noise reconstruction (Fig.~3(b,c) of the main text), we observe discrete noise features at frequencies \mbox{$\omega/2\pi \approx 5.2, 7.4, 13.5, 18.8$} and 31.4~kHz, corresponding to $\mathbb{E}[P_1]$ signal observed when performing square sequences with \mbox{$\tau = 2$~ms} and \mbox{$S=20,29,53,74 \text{ and }124$}. In the Slepian case, only two of these features are present, corresponding to $S=20$ and $S=124$ (see Fig.~\ref{fig:data}). The additional features observed using the square pulses may be attributed to either out-of-band signal picked up by the filter function harmonics, or amplitude noise related to the rapid switching of acousto-optic modulators. To investigate the origin of these features, we use a photodiode at the trap output to examine the amplitude noise spectrum of the bichromatic Raman beam, performing measurements with the laser continuously on, off and pulsed. We inspect the spectra measured by an amplified photodiode (Thorlabs PDA100A2) on a mixed signal oscilloscope and spectrum analyzer for features at the above frequencies, as well as for any other features to which the respective filter function main band or harmonics for a given sequence may be sensitive.

Shown in Fig.~\ref{fig:amplitude}(a) is the spectrum measured with the laser continuously on. The signal is from the bichromatic beam path, however, in this case the modulator is driven with a single tone, producing a continuous beam without any of the (desired) amplitude modulation at the beat note frequency of the two-tone drive. Present in the spectrum is a strong signal at $\sim 135$~kHz, which is a known modulation on the laser amplitude likely from an internal power-supply. In order to differentiate the laser signal from background electronic noise, we compare the spectrum with the laser off and on; the difference of these measurements (figure inset, red) indicates that all other spectral features are common-mode and appear to be due to electronic pickup on the photodiode or the cable used to perform the measurement.

Figure.~\ref{fig:amplitude}(b) shows a zoomed-in view of the photodiode spectra in the region from 1 to 50 kHz, where a number of discrete spectral features are present. We observe an approximate correspondence between the frequency of these features and the band of filter-function peak-frequency-sensitivity (gray shading) for the sequences listed above which resulted in $\mathbb{E}[P_1]$ signal. For the \mbox{$S=20$} and \mbox{$S=124$} sequences, this correspondence suggests that the $\sim 5.2$~kHz and $\sim 31.4$~kHz features measured in both the square and Slepian intrinsic-noise reconstruction may arise from pickup of these electronic signals on the trap electrodes or the resonator stabilisation circuit. Despite a similar correspondence between the $S=29$ sensitivity-band with an observed peak, we cannot attribute the $\sim 7.4$~kHz feature measured in the intrinsic-noise reconstruction to this electronic signal as it is only present in the square-pulse case.

\begin{figure*}
    \centering
    \includegraphics{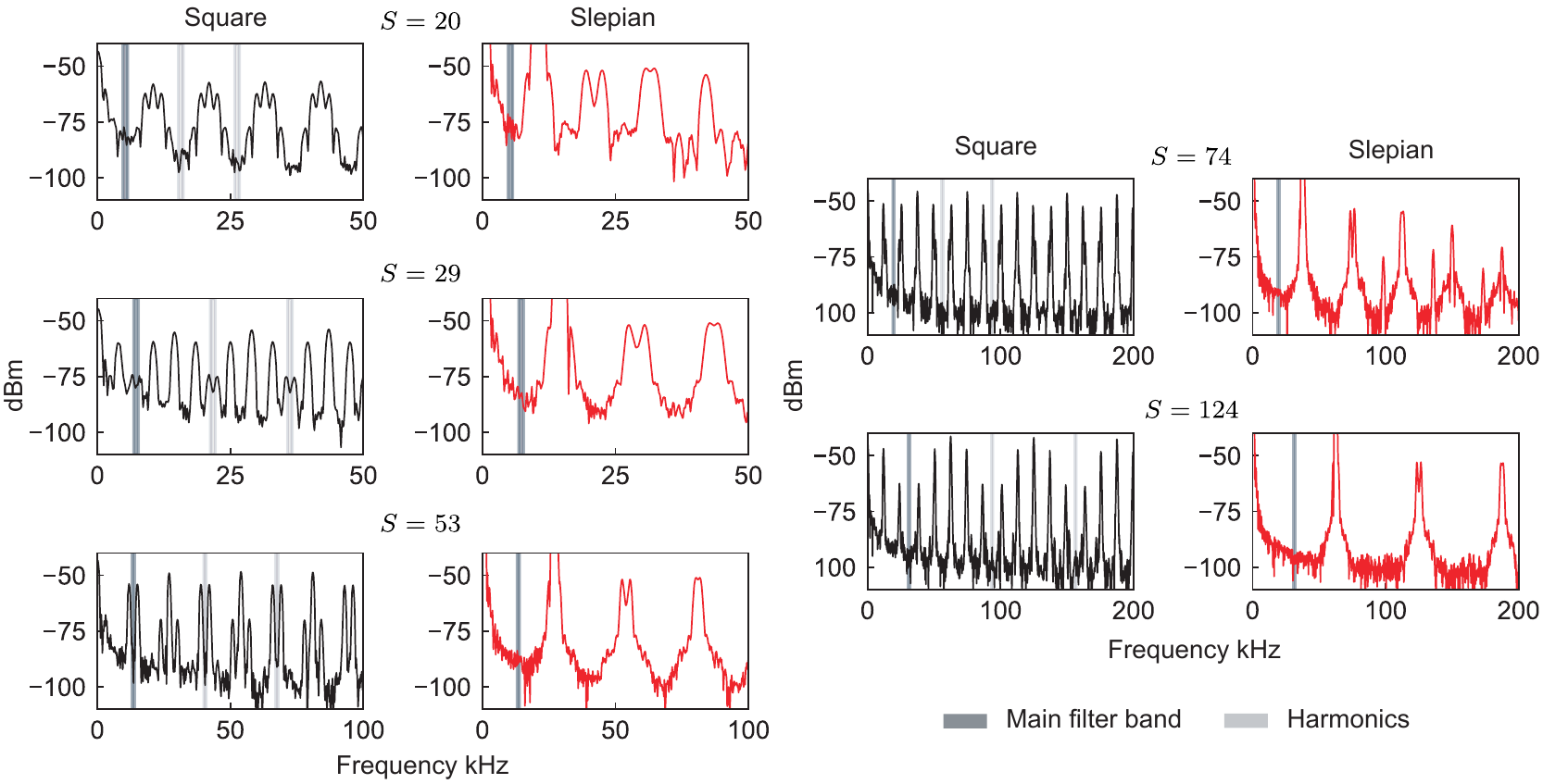}
    \caption{Fourier transform of the photodiode signal used to monitor the bichromatic 355~nm beam, measuring amplitude fluctuations of the pulsed sequences employed for noise sensing. Each row in the figure corresponds to a square (black) and Slepian (red) pulse for a fixed number of phase inversions, $S$. For square sequences, the harmonics are related to amplitude modulation caused by the discrete phase inversions, while the Slepians have a clear structure linked to even harmonics of the modulation frequency $\omega_{S}$.}
    \label{fig:FFTs}
\end{figure*}

To examine amplitude noise introduced by both the rf-phase-modulation employed in implementing the probe sequences and the switching of acousto-optic modulators, we also record amplitude noise spectra for the pulsed sequences (Fig.~\ref{fig:FFTs}). In this case, phase-modulated bichromatic laser pulses are incident on the photodiode. The photodiode bandwidth is limited to $\sim 260$~kHz, which filters the beat note modulation from the bichromatic light field at 3.2 MHz (twice the motional mode frequency).

In the case of square pulses, we observe multiple harmonics in the amplitude spectra relating to residual amplitude modulation at the locations of the discrete phase inversions used throughout the pulse. By contrast, for Slepian modulation, the phase inversions always occur at a point where the pulse amplitude goes to zero; accordingly we do not observe the same harmonic structure in the FFT. Instead, we observe harmonics at integer multiples of $2 \omega_S$, that is, twice the frequency of the underlying cosine modulation.

For the sequences with $S=20,74$ and 124, we do not observe overlap of the main filter band (or harmonics) with any significant spectral features. From this, we can likely exclude residual laser-amplitude modulation under sensing-protocol implementation as a potential source of the key features ($\sim 5.2$ and $\sim 31.4$ kHz) we observe in Fig.~3(b,c) of the main text. Additionally, the feature at $\sim 18.8$~kHz can likely be attributed to harmonic sensitivity outside of the main sensing band. The spectra for the $S=29$ and $S=53$ sequences, however, do reveal overlap between the filter band and the harmonics associated with phase inversion. This indicates that features at $\sim 7.4$ and $\sim 13.5$~kHz in the square-sequence intrinsic noise reconstruction may be due to amplitude noise, as opposed to motional frequency fluctuations - again demonstrating the advantage of deploying a Slepian pulse modulation scheme.

\clearpage